\documentclass{INTERSPEECH2023}

\usepackage{cite}
\usepackage{graphicx}  
\usepackage{float}  
\usepackage{subfigure}  
\usepackage{multirow}

\interspeechcameraready

\title{GenDistiller: Distilling Pre-trained Language Models based on Generative Models}
\name{Yingying Gao, Shilei Zhang, Zihao Cui, Yanhan Xu, Chao Deng, Junlan Feng$^{\ast}$ \thanks{*Corresponding author}}
\address{
  China Mobile Research}
\email{(gaoyingying,zhangshilei,cuizihaoyjy,dengchao,fengjunlan)@chinamobile.com}

\begin{document}

\maketitle
 
\begin{abstract}
Self-supervised pre-trained models such as HuBERT and WavLM leverage unlabeled speech data for representation learning and offer significantly improve for numerous downstream tasks. Despite the success of these methods, their large memory and strong computational requirements hinder their application on resource restricted devices. Therefore, this paper introduces GenDistiller, a novel knowledge distillation framework to distill hidden representations from teacher network based on generative language model. The generative structure enables the proposed model to generate the target teacher hidden layers autoregressively, considering the interactions between hidden layers without instroducing additional inputs. A two-dimensional attention mechanism is implemented to ensure the causality of hidden layers, while preserving bidirectional attention in the time dimension. Experiments reveal the advantage of the generative distiller over the baseline system that predicts the hidden layers of teacher network directly without a generatvie model.


\end{abstract}

\section{Introduction}

Pre-trained language models (PLMs), such as BERT \cite{bert}, RoBERTa \cite{roberta}, T5 \cite{t5}, wav2vec \cite{wav2vec}, WavLM \cite{wavlm}, HuBERT \cite{hubert} and speechT5 \cite{speecht5}, have been proven to bring significant improvement to a wide range of downstream tasks. However, PLMs usually suffer from their enormous parameters and long inference time, which obstruct their deployment in source-restricted applications. In addition, the transferring of PLMs together with a downstream model further exacerbates the consumption of computing and data resource. 

Pruning \cite{prune}, quantization \cite{quantization} and knowledge distillation (KD) \cite{kd} are the common techniques for model compression, aiming at reducing model size while retaining their performance as much as possible. In this paper, we focus on compressing PLMs based on KD, which trains a smaller student network under the supervision of a relatively large teacher network and presents significant advantages over direct training the student network using hard labels \cite{kd}.

Most of the effective KD methods \cite{distilbert,minilm,minilmv2,ernietiny,fitnets,deit} require a layer-to-layer distillation between intermediate layers rather than only the output layers, since different layers in PLMs contain different information and it is usually necessary to deliver multiple layers to downstream tasks as features to achieve better performance. This type of approaches need a certain correspondence between the architectures of student network and teacher network. Although some methods also attempt to relax \cite{tinybert} or automate \cite{dynabert} this correspondence, it is still not flexible enough. The work in \cite{distilhubert} breaks this restriction through appending several prediction heads on student network to predict the hidden layers’ output directly in teacher network. This simple framework provides a new paradigm for layer-wise knowledge distillation, without requiring a correspondence between the hidden layers of teacher network and student network. However, the prediction heads for different layers are implemented as multi-task learning which produce different hidden layers synchronously, and the interrelationships between them are not considered. Moreover, the synchronous prediction for multiple hidden layers results in a limited number of layers to be distilled, otherwise, the performance of certain downstream tasks will be compromised \cite{distilhubert}. We consider that the performance degradation is due to insufficient accuracy of hidden layer prediction. Therefore, we propose a new hidden layer prediction framework based on a generative language model. One reason for choosing a generative language model as the backbone of the distiller is that the generation process considers the interaction between the current target hidden layer and the previous layers of it, and the other reason is that the generative model does not introduce additional inputs other than the original features during the inference stage.

The major contributions of this work are three-fold:

1) We propose a knowledge distillation method based on generative models, which can achieve layer by layer distillation from pre-trained teacher models to smaller distillers.

2) We explore different attention mechanisms to implement two-dimensional attention, in order to achieve causal prediction for the hidden layer, while preserving bidirectional attention in the time dimension. 

3) This algorithm takes into account the correlation between the hidden layers of teacher model, and does not introduce additional inputs other than the original features during the inference stage.


\section{Related Work}

\begin{figure*}[h]
  \centering
  \includegraphics[width=\linewidth]{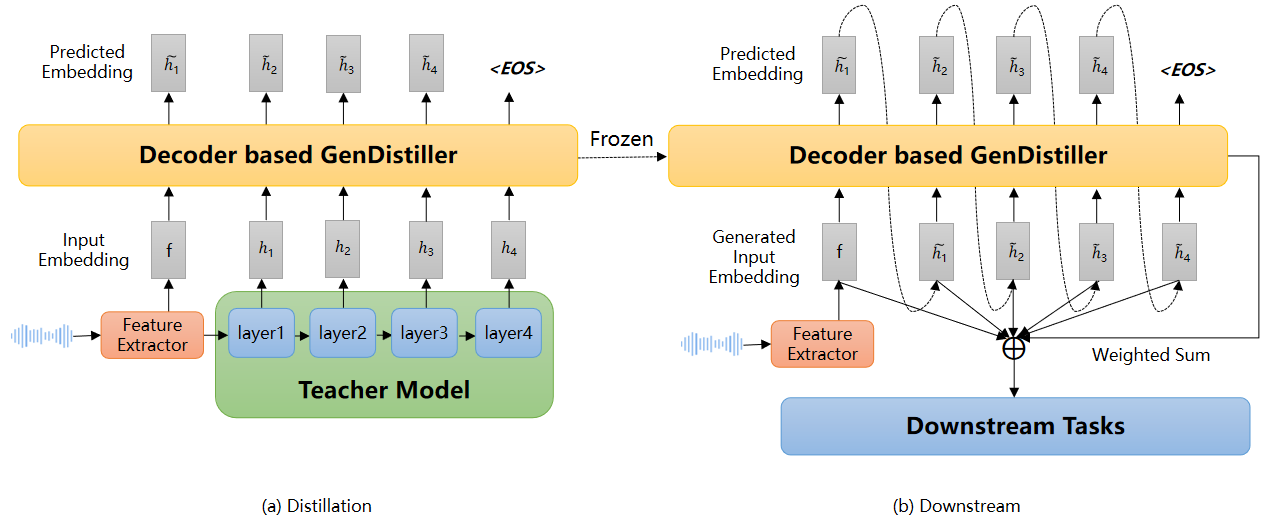}
  \caption{The distillation process and the downstream training of the proposed GenDistiller.}
  \label{fig1}
\end{figure*}

\subsection{Knowledge Distillation for PLMs}

Most previous work in KD \cite{tokd1,tokd2,tokd3} focus on task-oriented distillation while neglects some kind of task-agnostic knowledge, which plays a major role in the generality of downstream tasks. TinyBERT \cite{tinybert} deals with this problem through a two-stage learning framework, including the general distillation and the task-specific distillation, which is able to transfer generic knowledge from teacher to student network and achieves a better performance in specific downstream tasks. However, due to the significant reduction on model size, the TinyBERT before task-specific distillation performs generally worse than BERT. DistilBERT \cite{distilbert} is a distilled version of BERT with a general-purpose, which inherits 97\% of the teacher model’s capabilities but 40\% smaller and 60\% faster. It reduces the model size of BERT through reducing the number of layers by a factor of 2 and trains the distiller by a combined loss including the distillation loss and cosine-distance, as well as the language modeling loss. However, the layer number reduction method restricts the architecture of DistilBERT. DynaBERT \cite{dynabert} is a more flexible distiller of BERT which can adjust the width and depth to adapt different requirements for model size and latency. However, it still cannot completely avoid the layer-to-layer constraints between teacher and student networks. MINILM \cite{minilm} distills the self-attention module of the last Transformer layer which relaxes the layer mapping between teacher and student networks, and allows more flexibility for architecture of student model. MINILMv2 \cite{minilmv2} generalizes the self-attention distillation of MINILM by using multi-head self-attention relations computed by scaled dot-product of pairs of queries, keys and values, but these type of methods are only appropriate for models based on self-attention mechanism. Some hybrid methods \cite{structprune,lighthubert,dphubert} have integrated knowledge distillation and structure pruning and achieved very strong performances in Speech processing Universal PERformance Benchmark (SUPERB) \cite{superb}, but since pruning involves the searching for the sturcture of student network, this is not included in the research scope of this work. This work mainly focuses on whether generative architecture is more helpful for knowledge transfer from teachure network to student network, in other words, can large models directly generate small models, in a way of generating model outputs.

\subsection{Generative Language Models}

The language models can predict the embedding of the target token based on the context or the previous tokens of it. In this work, we use generative language model to predict the hidden layer outputs of teacher network in the hope to take the interaction of the hidden layers into account and avoid seeing the future information. Three language model architectures are considered: encoder-decoder, causal decoder, and prefix decoder \cite{survey}. The encoder-decoder architecture \cite{encdec,t5,bart} consists of two stacks of Transformer blocks as the encoder and decoder, in which encoder is to encode the input sequence as a common history for the generated sequence and decoder generates the target sequence autoregressively. The causal decoder architecture \cite{gpt1,gpt2,gpt3,gpt4} only attend to the past tokens of the input and itself through a unidirectional attention mask. The prefix decoder architecture \cite{palm,upalm} performs bidirectional attention over the prefix tokens and unidirectional attention only on generated tokens. In our work, we treat the original feature as the prefix tokens and the target hidden layers as the sequence to be generated one-by-one. Therefore, we select the prefix decoder architecture and modify it to some extend to build our distiller which can bidirectionally encode the input features and predict the output layers autoregressively, meanwhile, the required parameters are less than the encoder-decoder architecture.

\section{Method}

\begin{figure*}[h]
  \centering
  \includegraphics[width=\linewidth]{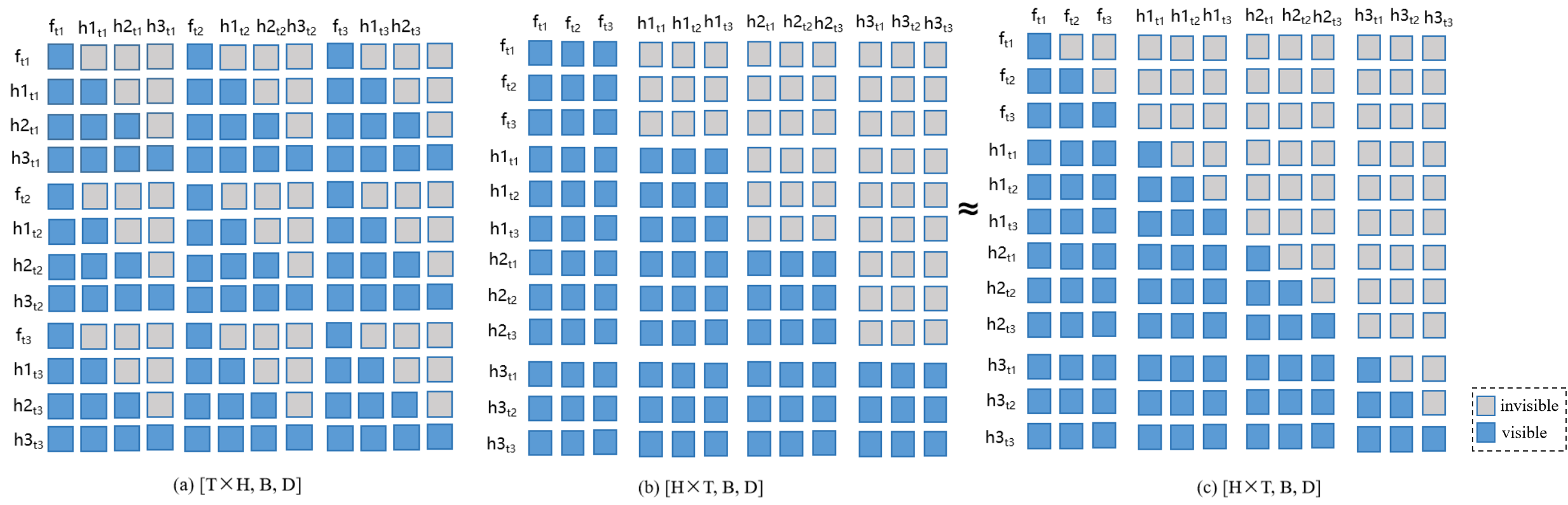}
  \caption{Different attention mechanisms via different flatten manners. a) flatten to $T\times H$ length for attention; b) flatten to $H\times T$ length for attention; c) the approximation of b).}
  \label{fig2}
\end{figure*}

We propose a knowledge distillation method based on generative models, which takes the correlation between the hidden layers of the teacher model into account and can implement layer by layer distillation from pre-trained teacher models to smaller distillers. The proposed framework are shown in Figure 1. During the distillation stage as shown in Figure~\ref{fig1}(a), the output embedding  of the hidden layers from the teacher model are set as the target sequence, which are delivered as input to the decoder after the original features. As mentioned above, we choose the modified prefix decoder architecture as the backbone of our distiller model, in which the input features are bidirectionally attended and the target hidden layers from the teacher network are predicted one by one seeing only the previous neighbors. It considers the interaction between the hidden layers and does not introduce additional inputs other than the original features during the inference stage. After distillation, the parameters of GenDistiller are forzen and the predicted layers generated by the distiller are appended one-by-one as the history to predict the next layer. The predicted hidden layers and the hidden layers of GenDistiller, as well as the original feature, are weighted summed and delivered to various downstream tasks as input features (shown in Figure~\ref{fig1}(b)).

\subsection{Two-dimensional Attention Mechanism}

The original feature  is a three-dimensional tensor, specifically, $f\in \mathbb{R} ^{T\times B\times D}$, in which $T$ is the frame numbers related with the sample length for self-attention, $B$ refers to the batch size and $D$ denotes the feature dimension. Originally, the input feature performs a bidirectional attention along the time dimension $T$ and the $B$ samples in one batch are processed in parallel. In this work, hidden layers are appended after feature $f$ and constitute a four-dimensional tensor, named as $M\in \mathbb{R} ^{H\times T\times B\times D}$, $H$ refers the numbers of hidden layers to be predicted plus the original feature. The time dimension permits bidirectional attention while the hidden layer dimension dose not, which means that the hidden layer is generated only based on the layers before it. 
   
If only the unidirectional attention is performed on the hidden layer dimension, which means the four-dimensional tensor $M$ is transformed into three-dimensional $[H, B\times T, D]$, then each frame $t_{i}$ is treated as a sample that will be processed in parallel with other frames in the batch, 
therefore the contextual impact of the time dimension will be ignored which has been proven to significantly decline the performance.

Instead, if the attention is implemented along the time dimension, different hidden layers are equivalent to different samples to be processed in parallel
. That is, the four-dimensional tensor $M$ is transformed into three-dimensional $[T, B\times H, D]$. All the $B\times H$ samples inside one batch are processed in parallel so as to avoid seeing the other layers around it. Meanwhile, since the prediction target of our model is to predict the next layer based on the current known layer, the interaction of adjacent layers is also involved. The deficiency is that it does not consider further historical information, especially the original features f. 

We tried to integrate these two attention methods. 
Two combinations manners are exploited, 1) performs unidirectional attention along hidden layer dimension first then attend bidirectionally along time dimension and 2) conversely. The results of them were both not satisfactory. 

Ultimately, we choose to flatten the two dimensions - time dimension and hidden layer dimension - to attend on. There are two possible flatten manners, as shown in Figure~\ref{fig2}(a) and (b), flatten into $T\times H$ length or $H\times T$ length. Although the final length is equal, the areas that can be seen during self attention are different due to the different sorting of the hidden layer dimension and the time dimension. As we can see in Figure~\ref{fig2}(a), the visible and invisible areas are scattered, and the overall trend does not exhibit  due to the causality of hidden layer dimension. On the contrary, if we place the hidden layer in the outermost dimension, each small region within the interior is continuously visible or invisible, and the overall causal trend can also be observed. In the end, we only need to sacrifice a few points to achieve the approximation of Figure~\ref{fig2}(b) with a simple global unidirectional attention, as shown in Figure~\ref{fig2}(c).

\subsection{Training}

During distillation stage, the parameters of teacher model are frozen and only the GenDistiller are updated. The output embedding of hidden layers in teacher model are delivered in parallell to GenDistiller together with the original feature, while the predicted layers generated by the distiller are appended one-by-one as the history to predict the next layer during the inference stage. To alleviate inconsistencies between the two stages, we incorporate the two types of input manners in the later stage of training for a better performance.

\subsection{Distillation Loss}

To compare with other work, the distillation loss follows the loss function of DistilHuBERT \cite{distilhubert}, which consists of two parts: $\ell_{1}$ distance $\mathcal{L}_{\ell_{1}}$ and cosine similarity $\mathcal{L}_{cos}$ between the $l_{th}$ target hidden layer $h_{t}^{(l)} $ at time $t_{th}$ from teacher model and the generated layer $\widetilde{h}_{t}^{(l)}$ produced by student model. The loss functions is



\begin{equation}
  \begin{split}
    & \mathcal{L}^{(l)}=\mathcal{L}_{\ell_{1}}^{(l)}+\lambda\mathcal{L}_{cos}^{(l)} \\
    & =\sum_{t=1}^{T}\left [\frac{1}{D}\Vert \widetilde{h}_{t}^{(l)}-h_{t}^{(l)} \Vert_{1}-\lambda \log{\sigma (\cos (\widetilde{h}_{t}^{(l)},h_{t}^{(l)}))}\right ] \\
  \end{split}
\end{equation}

Where $T$ refers to the time steps. $\sigma$ is sigmoid activation and  $\lambda$ controls the contribution of the cosine similarity loss.

\section{Experiments}
\subsection{Experimental Setup}
We implement our experiments with S3PRL \cite{s3prl1,s3prl2}. The proposed model is evaluated on automatic speech recognition (ASR) in SUPERB \cite{superb}. 

\noindent\textbf{Data}.  All the 960 hours of training data in LibriSpeech \cite{librispeech} are used for knowledge distillation except the experiment used AISHELL-1 \cite{aishell1} to see the effect of different data sets. The training of ASR task is carried out by 100 hours of clean training data in LibriSpeech. 

\noindent\textbf{Model}. We choose the WavLM base model as the teacher model, which consists of a 7-layer CNN feature extractor and a 12-layer transformer encoder. Our GenDistiller has a similar architecture with DistilHuBERT in order to compare their performances. It is constituted by two transformer layers as same as the ones in WavLM base model. And the linear output layer after the two transformer layers in DistilHuBERT is also remained.

\noindent\textbf{Training}. The distillation procedure is run on a 32GB V100 GPU for 200k steps with a batch size of 24 utterances. The training hyper parameters are setting according to DistilHuBERT. The ASR downstream task is trained for 200k steps with a batch size of 32 utterances.

\subsection{Results}

We first compare the the performance of the proposed model and baselines. The mean values of the two hidden layers of the distiller for different inputs are taken as pretrained feature for downstream tasks. The model is trained to predict the $4^{th}$, $8^{th}$ and $12^{th}$ layers of the teacher model since it is tested to be the best selection for DistilHuBERT. Table~\ref{tab1} shows that the proposed GenDisilWavLM exceed the other distilled models without using generative models as distiller.

More experiments are underway and will be reported in the future.


\begin{table}[th]
  \caption{Results on SUPERB of the proposal and baselines. The proposed model is evaluated on automatic speech recognition (ASR) is measured with word error rate (WER\%).} 
  \label{tab1}
  \centering
  \begin{tabular}{l l}
    \hline
    \multirow{2}*{\textbf{Method}} & \textbf{ASR}\\
    \cline{2-2}
    ~ & WER$\downarrow$ \\
    \hline
    \textbf{Baselines} & \\
    \hline
    HuBERT Base       & $6.42$     \\  
    WavLM Base       & $6.21$     \\ 
    \hline
    \textbf{Distilled Models} & \\
    DistilHuBERT  &$13.37$  \\
    DisilWavLM  & $13.24$  \\
    \hline
    \textbf{Ours} & \\
    GenDisilWavLM  &$12.53$  \\
    \hline
  \end{tabular} 
\end{table}

\section{Conclusion}
In this work, we propose a generative knowledge distilling model for large-scale pretrained language models. The proposed GenDistiller is able to generate the target teacher hidden layers autoregressively, taking the interactions between hidden layers into account and does not introduce additional inputs other than the original features. Different attention mechanisms are explored to ensure the causality of hidden layers, while preserving bidirectional attention in the time dimension. The final proposal surpasses the baselines that predict the hidden layers of teacher network directly without a generatvie model.


\bibliographystyle{IEEEtran}
\bibliography{mybib}

\begin{thebibliography}{10}
\providecommand{\url}[1]{#1}
\csname url@samestyle\endcsname
\providecommand{\newblock}{\relax}
\providecommand{\bibinfo}[2]{#2}
\providecommand{\BIBentrySTDinterwordspacing}{\spaceskip=0pt\relax}
\providecommand{\BIBentryALTinterwordstretchfactor}{4}
\providecommand{\BIBentryALTinterwordspacing}{\spaceskip=\fontdimen2\font plus
\BIBentryALTinterwordstretchfactor\fontdimen3\font minus
  \fontdimen4\font\relax}
\providecommand{\BIBforeignlanguage}[2]{{%
\expandafter\ifx\csname l@#1\endcsname\relax
\typeout{** WARNING: IEEEtran.bst: No hyphenation pattern has been}%
\typeout{** loaded for the language `#1'. Using the pattern for}%
\typeout{** the default language instead.}%
\else
\language=\csname l@#1\endcsname
\fi
#2}}
\providecommand{\BIBdecl}{\relax}
\BIBdecl

\bibitem{bert}
\BIBentryALTinterwordspacing
J.~Devlin, M.~Chang, K.~Lee, and K.~Toutanova, ``{BERT:} pre-training of deep
  bidirectional transformers for language understanding,'' in \emph{Proceedings
  of the 2019 Conference of the North American Chapter of the Association for
  Computational Linguistics: Human Language Technologies, {NAACL-HLT} 2019,
  Minneapolis, MN, USA, June 2-7, 2019, Volume 1 (Long and Short Papers)},
  J.~Burstein, C.~Doran, and T.~Solorio, Eds.\hskip 1em plus 0.5em minus
  0.4em\relax Association for Computational Linguistics, 2019, pp. 4171--4186.
  [Online]. Available: \url{https://doi.org/10.18653/v1/n19-1423}
\BIBentrySTDinterwordspacing

\bibitem{roberta}
\BIBentryALTinterwordspacing
Y.~Liu, M.~Ott, N.~Goyal, J.~Du, M.~Joshi, D.~Chen, O.~Levy, M.~Lewis,
  L.~Zettlemoyer, and V.~Stoyanov, ``Roberta: {A} robustly optimized {BERT}
  pretraining approach,'' \emph{CoRR}, vol. abs/1907.11692, 2019. [Online].
  Available: \url{http://arxiv.org/abs/1907.11692}
\BIBentrySTDinterwordspacing

\bibitem{t5}
\BIBentryALTinterwordspacing
C.~Raffel, N.~Shazeer, A.~Roberts, K.~Lee, S.~Narang, M.~Matena, Y.~Zhou,
  W.~Li, and P.~J. Liu, ``Exploring the limits of transfer learning with a
  unified text-to-text transformer,'' \emph{J. Mach. Learn. Res.}, vol.~21, pp.
  140:1--140:67, 2020. [Online]. Available:
  \url{http://jmlr.org/papers/v21/20-074.html}
\BIBentrySTDinterwordspacing

\bibitem{wav2vec}
S.~Schneider, A.~Baevski, R.~Collobert, and M.~Auli, ``{wav2vec: Unsupervised
  Pre-Training for Speech Recognition},'' in \emph{Proc. Interspeech 2019},
  2019, pp. 3465--3469.

\bibitem{wavlm}
\BIBentryALTinterwordspacing
S.~Chen, C.~Wang, Z.~Chen, Y.~Wu, S.~Liu, Z.~Chen, J.~Li, N.~Kanda,
  T.~Yoshioka, X.~Xiao, J.~Wu, L.~Zhou, S.~Ren, Y.~Qian, Y.~Qian, J.~Wu,
  M.~Zeng, X.~Yu, and F.~Wei, ``Wavlm: Large-scale self-supervised pre-training
  for full stack speech processing,'' \emph{{IEEE} J. Sel. Top. Signal
  Process.}, vol.~16, no.~6, pp. 1505--1518, 2022. [Online]. Available:
  \url{https://doi.org/10.1109/JSTSP.2022.3188113}
\BIBentrySTDinterwordspacing

\bibitem{hubert}
\BIBentryALTinterwordspacing
W.~Hsu, B.~Bolte, Y.~H. Tsai, K.~Lakhotia, R.~Salakhutdinov, and A.~Mohamed,
  ``Hubert: Self-supervised speech representation learning by masked prediction
  of hidden units,'' \emph{{IEEE} {ACM} Trans. Audio Speech Lang. Process.},
  vol.~29, pp. 3451--3460, 2021. [Online]. Available:
  \url{https://doi.org/10.1109/TASLP.2021.3122291}
\BIBentrySTDinterwordspacing

\bibitem{speecht5}
\BIBentryALTinterwordspacing
J.~Ao, R.~Wang, L.~Zhou, C.~Wang, S.~Ren, Y.~Wu, S.~Liu, T.~Ko, Q.~Li,
  Y.~Zhang, Z.~Wei, Y.~Qian, J.~Li, and F.~Wei, ``Speecht5: Unified-modal
  encoder-decoder pre-training for spoken language processing,'' in
  \emph{Proceedings of the 60th Annual Meeting of the Association for
  Computational Linguistics (Volume 1: Long Papers), {ACL} 2022, Dublin,
  Ireland, May 22-27, 2022}, S.~Muresan, P.~Nakov, and A.~Villavicencio,
  Eds.\hskip 1em plus 0.5em minus 0.4em\relax Association for Computational
  Linguistics, 2022, pp. 5723--5738. [Online]. Available:
  \url{https://doi.org/10.18653/v1/2022.acl-long.393}
\BIBentrySTDinterwordspacing

\bibitem{prune}
\BIBentryALTinterwordspacing
S.~Han, J.~Pool, J.~Tran, and W.~J. Dally, ``Learning both weights and
  connections for efficient neural network,'' in \emph{Neural Information
  Processing Systems}, 2015. [Online]. Available:
  \url{https://api.semanticscholar.org/CorpusID:2238772}
\BIBentrySTDinterwordspacing

\bibitem{quantization}
\BIBentryALTinterwordspacing
Y.~Gong, L.~Liu, M.~Yang, and L.~D. Bourdev, ``Compressing deep convolutional
  networks using vector quantization,'' \emph{CoRR}, vol. abs/1412.6115, 2014.
  [Online]. Available: \url{http://arxiv.org/abs/1412.6115}
\BIBentrySTDinterwordspacing

\bibitem{kd}
\BIBentryALTinterwordspacing
G.~E. Hinton, O.~Vinyals, and J.~Dean, ``Distilling the knowledge in a neural
  network,'' \emph{CoRR}, vol. abs/1503.02531, 2015. [Online]. Available:
  \url{http://arxiv.org/abs/1503.02531}
\BIBentrySTDinterwordspacing

\bibitem{distilbert}
\BIBentryALTinterwordspacing
V.~Sanh, L.~Debut, J.~Chaumond, and T.~Wolf, ``Distilbert, a distilled version
  of {BERT:} smaller, faster, cheaper and lighter,'' \emph{CoRR}, vol.
  abs/1910.01108, 2019. [Online]. Available:
  \url{http://arxiv.org/abs/1910.01108}
\BIBentrySTDinterwordspacing

\bibitem{minilm}
\BIBentryALTinterwordspacing
W.~Wang, F.~Wei, L.~Dong, H.~Bao, N.~Yang, and M.~Zhou, ``Minilm: Deep
  self-attention distillation for task-agnostic compression of pre-trained
  transformers,'' in \emph{Advances in Neural Information Processing Systems
  33: Annual Conference on Neural Information Processing Systems 2020, NeurIPS
  2020, December 6-12, 2020, virtual}, H.~Larochelle, M.~Ranzato, R.~Hadsell,
  M.~Balcan, and H.~Lin, Eds., 2020. [Online]. Available:
  \url{https://proceedings.neurips.cc/paper/2020/hash/3f5ee243547dee91fbd053c1c4a845aa-Abstract.html}
\BIBentrySTDinterwordspacing

\bibitem{minilmv2}
\BIBentryALTinterwordspacing
W.~Wang, H.~Bao, S.~Huang, L.~Dong, and F.~Wei, ``Minilmv2: Multi-head
  self-attention relation distillation for compressing pretrained
  transformers,'' in \emph{Findings of the Association for Computational
  Linguistics: {ACL/IJCNLP} 2021, Online Event, August 1-6, 2021}, ser.
  Findings of {ACL}, C.~Zong, F.~Xia, W.~Li, and R.~Navigli, Eds., vol.
  {ACL/IJCNLP} 2021.\hskip 1em plus 0.5em minus 0.4em\relax Association for
  Computational Linguistics, 2021, pp. 2140--2151. [Online]. Available:
  \url{https://doi.org/10.18653/v1/2021.findings-acl.188}
\BIBentrySTDinterwordspacing

\bibitem{ernietiny}
\BIBentryALTinterwordspacing
W.~Su, X.~Chen, S.~Feng, J.~Liu, W.~Liu, Y.~Sun, H.~Tian, H.~Wu, and H.~Wang,
  ``Ernie-tiny : {A} progressive distillation framework for pretrained
  transformer compression,'' \emph{CoRR}, vol. abs/2106.02241, 2021. [Online].
  Available: \url{https://arxiv.org/abs/2106.02241}
\BIBentrySTDinterwordspacing

\bibitem{fitnets}
\BIBentryALTinterwordspacing
A.~Romero, N.~Ballas, S.~E. Kahou, A.~Chassang, C.~Gatta, and Y.~Bengio,
  ``Fitnets: Hints for thin deep nets,'' in \emph{3rd International Conference
  on Learning Representations, {ICLR} 2015, San Diego, CA, USA, May 7-9, 2015,
  Conference Track Proceedings}, Y.~Bengio and Y.~LeCun, Eds., 2015. [Online].
  Available: \url{http://arxiv.org/abs/1412.6550}
\BIBentrySTDinterwordspacing

\bibitem{deit}
\BIBentryALTinterwordspacing
H.~Touvron, M.~Cord, M.~Douze, F.~Massa, A.~Sablayrolles, and H.~J{\'{e}}gou,
  ``Training data-efficient image transformers {\&} distillation through
  attention,'' in \emph{Proceedings of the 38th International Conference on
  Machine Learning, {ICML} 2021, 18-24 July 2021, Virtual Event}, ser.
  Proceedings of Machine Learning Research, M.~Meila and T.~Zhang, Eds., vol.
  139.\hskip 1em plus 0.5em minus 0.4em\relax {PMLR}, 2021, pp.
  10\,347--10\,357. [Online]. Available:
  \url{http://proceedings.mlr.press/v139/touvron21a.html}
\BIBentrySTDinterwordspacing

\bibitem{tinybert}
\BIBentryALTinterwordspacing
X.~Jiao, Y.~Yin, L.~Shang, X.~Jiang, X.~Chen, L.~Li, F.~Wang, and Q.~Liu,
  ``Tinybert: Distilling {BERT} for natural language understanding,'' in
  \emph{Findings of the Association for Computational Linguistics: {EMNLP}
  2020, Online Event, 16-20 November 2020}, ser. Findings of {ACL}, T.~Cohn,
  Y.~He, and Y.~Liu, Eds., vol. {EMNLP} 2020.\hskip 1em plus 0.5em minus
  0.4em\relax Association for Computational Linguistics, 2020, pp. 4163--4174.
  [Online]. Available:
  \url{https://doi.org/10.18653/v1/2020.findings-emnlp.372}
\BIBentrySTDinterwordspacing

\bibitem{dynabert}
\BIBentryALTinterwordspacing
L.~Hou, Z.~Huang, L.~Shang, X.~Jiang, X.~Chen, and Q.~Liu, ``Dynabert: Dynamic
  {BERT} with adaptive width and depth,'' in \emph{Advances in Neural
  Information Processing Systems 33: Annual Conference on Neural Information
  Processing Systems 2020, NeurIPS 2020, December 6-12, 2020, virtual},
  H.~Larochelle, M.~Ranzato, R.~Hadsell, M.~Balcan, and H.~Lin, Eds., 2020.
  [Online]. Available:
  \url{https://proceedings.neurips.cc/paper/2020/hash/6f5216f8d89b086c18298e043bfe48ed-Abstract.html}
\BIBentrySTDinterwordspacing

\bibitem{distilhubert}
\BIBentryALTinterwordspacing
H.~Chang, S.~Yang, and H.~Lee, ``Distilhubert: Speech representation learning
  by layer-wise distillation of hidden-unit bert,'' in \emph{{IEEE}
  International Conference on Acoustics, Speech and Signal Processing, {ICASSP}
  2022, Virtual and Singapore, 23-27 May 2022}.\hskip 1em plus 0.5em minus
  0.4em\relax {IEEE}, 2022, pp. 7087--7091. [Online]. Available:
  \url{https://doi.org/10.1109/ICASSP43922.2022.9747490}
\BIBentrySTDinterwordspacing

\bibitem{tokd1}
\BIBentryALTinterwordspacing
L.~Zhang, Y.~Shi, Z.~Shi, K.~Ma, and C.~Bao, ``Task-oriented feature
  distillation,'' in \emph{Advances in Neural Information Processing Systems
  33: Annual Conference on Neural Information Processing Systems 2020, NeurIPS
  2020, December 6-12, 2020, virtual}, H.~Larochelle, M.~Ranzato, R.~Hadsell,
  M.~Balcan, and H.~Lin, Eds., 2020. [Online]. Available:
  \url{https://proceedings.neurips.cc/paper/2020/hash/a96b65a721e561e1e3de768ac819ffbb-Abstract.html}
\BIBentrySTDinterwordspacing

\bibitem{tokd2}
\BIBentryALTinterwordspacing
W.~Huang, Z.~Peng, L.~Dong, F.~Wei, J.~Jiao, and Q.~Ye, ``Generic-to-specific
  distillation of masked autoencoders,'' in \emph{{IEEE/CVF} Conference on
  Computer Vision and Pattern Recognition, {CVPR} 2023, Vancouver, BC, Canada,
  June 17-24, 2023}.\hskip 1em plus 0.5em minus 0.4em\relax {IEEE}, 2023, pp.
  15\,996--16\,005. [Online]. Available:
  \url{https://doi.org/10.1109/CVPR52729.2023.01535}
\BIBentrySTDinterwordspacing

\bibitem{tokd3}
\BIBentryALTinterwordspacing
M.~Xue, J.~Song, X.~Wang, Y.~Chen, X.~Wang, and M.~Song, ``Kdexplainer: {A}
  task-oriented attention model for explaining knowledge distillation,'' in
  \emph{Proceedings of the Thirtieth International Joint Conference on
  Artificial Intelligence, {IJCAI} 2021, Virtual Event / Montreal, Canada,
  19-27 August 2021}, Z.~Zhou, Ed.\hskip 1em plus 0.5em minus 0.4em\relax
  ijcai.org, 2021, pp. 3228--3234. [Online]. Available:
  \url{https://doi.org/10.24963/ijcai.2021/444}
\BIBentrySTDinterwordspacing

\bibitem{structprune}
H.~Wang, S.~Wang, W.-Q. Zhang, S.~Hongbin, and Y.~Wan, ``{Task-Agnostic
  Structured Pruning of Speech Representation Models},'' in \emph{Proc.
  INTERSPEECH 2023}, 2023, pp. 231--235.

\bibitem{lighthubert}
R.~Wang, Q.~Bai, J.~Ao, L.~Zhou, Z.~Xiong, Z.~Wei, Y.~Zhang, T.~Ko, and H.~Li,
  ``{LightHuBERT: Lightweight and Configurable Speech Representation Learning
  with Once-for-All Hidden-Unit BERT},'' in \emph{Proc. Interspeech 2022},
  2022, pp. 1686--1690.

\bibitem{dphubert}
Y.~Peng, Y.~Sudo, S.~Muhammad, and S.~Watanabe, ``{DPHuBERT: Joint Distillation
  and Pruning of Self-Supervised Speech Models},'' in \emph{Proc. INTERSPEECH
  2023}, 2023, pp. 62--66.

\bibitem{superb}
S.~wen Yang, P.-H. Chi, Y.-S. Chuang, C.-I.~J. Lai, K.~Lakhotia, Y.~Y. Lin,
  A.~T. Liu, J.~Shi, X.~Chang, G.-T. Lin, T.-H. Huang, W.-C. Tseng, K.~tik Lee,
  D.-R. Liu, Z.~Huang, S.~Dong, S.-W. Li, S.~Watanabe, A.~Mohamed, and
  H.~yi~Lee, ``{SUPERB: Speech Processing Universal PERformance Benchmark},''
  in \emph{Proc. Interspeech 2021}, 2021, pp. 1194--1198.

\bibitem{survey}
\BIBentryALTinterwordspacing
W.~X. Zhao, K.~Zhou, J.~Li, T.~Tang, X.~Wang, Y.~Hou, Y.~Min, B.~Zhang,
  J.~Zhang, Z.~Dong, Y.~Du, C.~Yang, Y.~Chen, Z.~Chen, J.~Jiang, R.~Ren, Y.~Li,
  X.~Tang, Z.~Liu, P.~Liu, J.~Nie, and J.~Wen, ``A survey of large language
  models,'' \emph{CoRR}, vol. abs/2303.18223, 2023. [Online]. Available:
  \url{https://doi.org/10.48550/arXiv.2303.18223}
\BIBentrySTDinterwordspacing

\bibitem{encdec}
\BIBentryALTinterwordspacing
A.~Vaswani, N.~Shazeer, N.~Parmar, J.~Uszkoreit, L.~Jones, A.~N. Gomez,
  L.~Kaiser, and I.~Polosukhin, ``Attention is all you need,'' in
  \emph{Advances in Neural Information Processing Systems 30: Annual Conference
  on Neural Information Processing Systems 2017, December 4-9, 2017, Long
  Beach, CA, {USA}}, I.~Guyon, U.~von Luxburg, S.~Bengio, H.~M. Wallach,
  R.~Fergus, S.~V.~N. Vishwanathan, and R.~Garnett, Eds., 2017, pp. 5998--6008.
  [Online]. Available:
  \url{https://proceedings.neurips.cc/paper/2017/hash/3f5ee243547dee91fbd053c1c4a845aa-Abstract.html}
\BIBentrySTDinterwordspacing

\bibitem{bart}
\BIBentryALTinterwordspacing
M.~Lewis, Y.~Liu, N.~Goyal, M.~Ghazvininejad, A.~Mohamed, O.~Levy, V.~Stoyanov,
  and L.~Zettlemoyer, ``{BART:} denoising sequence-to-sequence pre-training for
  natural language generation, translation, and comprehension,'' in
  \emph{Proceedings of the 58th Annual Meeting of the Association for
  Computational Linguistics, {ACL} 2020, Online, July 5-10, 2020}, D.~Jurafsky,
  J.~Chai, N.~Schluter, and J.~R. Tetreault, Eds.\hskip 1em plus 0.5em minus
  0.4em\relax Association for Computational Linguistics, 2020, pp. 7871--7880.
  [Online]. Available: \url{https://doi.org/10.18653/v1/2020.acl-main.703}
\BIBentrySTDinterwordspacing

\bibitem{gpt1}
\BIBentryALTinterwordspacing
A.~Radford and K.~Narasimhan, ``Improving language understanding by generative
  pre-training,'' 2018. [Online]. Available:
  \url{https://api.semanticscholar.org/CorpusID:49313245}
\BIBentrySTDinterwordspacing

\bibitem{gpt2}
\BIBentryALTinterwordspacing
A.~Radford, J.~Wu, R.~Child, D.~Luan, D.~Amodei, and I.~Sutskever, ``Language
  models are unsupervised multitask learners,'' 2019. [Online]. Available:
  \url{https://api.semanticscholar.org/CorpusID:160025533}
\BIBentrySTDinterwordspacing

\bibitem{gpt3}
\BIBentryALTinterwordspacing
T.~B. Brown, B.~Mann, N.~Ryder, M.~Subbiah, J.~Kaplan, P.~Dhariwal,
  A.~Neelakantan, P.~Shyam, G.~Sastry, A.~Askell, S.~Agarwal,
  A.~Herbert{-}Voss, G.~Krueger, T.~Henighan, R.~Child, A.~Ramesh, D.~M.
  Ziegler, J.~Wu, C.~Winter, C.~Hesse, M.~Chen, E.~Sigler, M.~Litwin, S.~Gray,
  B.~Chess, J.~Clark, C.~Berner, S.~McCandlish, A.~Radford, I.~Sutskever, and
  D.~Amodei, ``Language models are few-shot learners,'' in \emph{Advances in
  Neural Information Processing Systems 33: Annual Conference on Neural
  Information Processing Systems 2020, NeurIPS 2020, December 6-12, 2020,
  virtual}, H.~Larochelle, M.~Ranzato, R.~Hadsell, M.~Balcan, and H.~Lin, Eds.,
  2020. [Online]. Available:
  \url{https://proceedings.neurips.cc/paper/2020/hash/1457c0d6bfcb4967418bfb8ac142f64a-Abstract.html}
\BIBentrySTDinterwordspacing

\bibitem{gpt4}
\BIBentryALTinterwordspacing
OpenAI, ``{GPT-4} technical report,'' \emph{CoRR}, vol. abs/2303.08774, 2023.
  [Online]. Available: \url{https://doi.org/10.48550/arXiv.2303.08774}
\BIBentrySTDinterwordspacing

\bibitem{palm}
\BIBentryALTinterwordspacing
A.~Chowdhery, S.~Narang, J.~Devlin, M.~Bosma, G.~Mishra, A.~Roberts, P.~Barham,
  H.~W. Chung, C.~Sutton, S.~Gehrmann, P.~Schuh, K.~Shi, S.~Tsvyashchenko,
  J.~Maynez, A.~Rao, P.~Barnes, Y.~Tay, N.~Shazeer, V.~Prabhakaran, E.~Reif,
  N.~Du, B.~Hutchinson, R.~Pope, J.~Bradbury, J.~Austin, M.~Isard,
  G.~Gur{-}Ari, P.~Yin, T.~Duke, A.~Levskaya, S.~Ghemawat, S.~Dev,
  H.~Michalewski, X.~Garcia, V.~Misra, K.~Robinson, L.~Fedus, D.~Zhou,
  D.~Ippolito, D.~Luan, H.~Lim, B.~Zoph, A.~Spiridonov, R.~Sepassi, D.~Dohan,
  S.~Agrawal, M.~Omernick, A.~M. Dai, T.~S. Pillai, M.~Pellat, A.~Lewkowycz,
  E.~Moreira, R.~Child, O.~Polozov, K.~Lee, Z.~Zhou, X.~Wang, B.~Saeta,
  M.~Diaz, O.~Firat, M.~Catasta, J.~Wei, K.~Meier{-}Hellstern, D.~Eck, J.~Dean,
  S.~Petrov, and N.~Fiedel, ``Palm: Scaling language modeling with pathways,''
  \emph{CoRR}, vol. abs/2204.02311, 2022. [Online]. Available:
  \url{https://doi.org/10.48550/arXiv.2204.02311}
\BIBentrySTDinterwordspacing

\bibitem{upalm}
\BIBentryALTinterwordspacing
Y.~Tay, J.~Wei, H.~W. Chung, V.~Q. Tran, D.~R. So, S.~Shakeri, X.~Garcia, H.~S.
  Zheng, J.~Rao, A.~Chowdhery, D.~Zhou, D.~Metzler, S.~Petrov, N.~Houlsby,
  Q.~V. Le, and M.~Dehghani, ``Transcending scaling laws with 0.1{\%} extra
  compute,'' \emph{CoRR}, vol. abs/2210.11399, 2022. [Online]. Available:
  \url{https://doi.org/10.48550/arXiv.2210.11399}
\BIBentrySTDinterwordspacing

\bibitem{s3prl1}
A.~T. Liu, S.-W. Li, and H.~yi~Lee, ``Tera: Self-supervised learning of
  transformer encoder representation for speech,'' 2020.

\bibitem{s3prl2}
\BIBentryALTinterwordspacing
A.~T. Liu, S.-w. Yang, P.-H. Chi, P.-c. Hsu, and H.-y. Lee, ``Mockingjay:
  Unsupervised speech representation learning with deep bidirectional
  transformer encoders,'' \emph{ICASSP 2020 - 2020 IEEE International
  Conference on Acoustics, Speech and Signal Processing (ICASSP)}, May 2020.
  [Online]. Available: \url{http://dx.doi.org/10.1109/ICASSP40776.2020.9054458}
\BIBentrySTDinterwordspacing

\bibitem{librispeech}
V.~Panayotov, G.~Chen, D.~Povey, and S.~Khudanpur, ``Librispeech: an asr corpus
  based on public domain audio books,'' in \emph{Acoustics, Speech and Signal
  Processing (ICASSP), 2015 IEEE International Conference on}.\hskip 1em plus
  0.5em minus 0.4em\relax IEEE, 2015, pp. 5206--5210.

\bibitem{aishell1}
X.~N. B. W. H.~Z. Hui~Bu, Jiayu~Du, ``Aishell-1: An open-source mandarin speech
  corpus and a speech recognition baseline,'' in \emph{Oriental COCOSDA 2017},
  2017, p. Submitted.

\end{thebibliography}

\end{document}